\documentclass{natureASTRO}

\usepackage{pdfpages}

\usepackage{amsmath}	
\usepackage{graphicx}
\usepackage{amssymb}
\usepackage{chemformula}
\newcommand{\memsai}{{\it Mem.\,Soc.\,Astron.\,It.\, }}
\newcommand{\apj}{{\it Astrophys. J.\, }}
\newcommand{\pasp}{{\it  Publ. Astron. Soc. Pac.\, }}
\newcommand{\aapr}{{\it  Astron. Astrophys. Rev.\, }}
\newcommand{\actaa}{{\it  Acta Astron.\, }}
\newcommand{\aj}{{\it  Astronomical. J.\, }}
\newcommand{\apjl}{{\it  Astrophys. J. Lett.\, }}
\newcommand{\apjs}{{\it  Astrophys. J. Suppl.\, }}
\newcommand{\apss}{{\it  Publ. Astron. Soc. Pac.\, }}
\newcommand{\mnras}{{\it  Mon. Not. R. Astron. Soc.\, }}
\newcommand{\nat}{{\it  Nat.\, }}
\newcommand{\natast}{{\it  Nat. Ast.\, }}

\newcommand{\aap}{{\it  Astron. Astrophys.\, }}

\newcommand{\procspie}{{\it  Proc. of SPIE\, }}

\newcommand{\astrep}{{\it  Astron. Rep.\, }}

\usepackage{lineno}
\usepackage{rotating}
\usepackage{arydshln}
\title{A Plague of Magnetic Spots Among the Hot Stars of Globular Clusters}
\author{Y.     Momany$^{1}$,  S.     Zaggia$^{1}$,     
M. Montalto$^{2}$,  D.     Jones$^{3,4}$,   H.M.J. Boffin$^{5}$, 
S. Cassisi$^{6,7}$, C. Moni Bidin$^{8}$,  M.   Gullieuszik$^{1}$, 
I.     Saviane$^{9}$,  L.    Monaco$^{10}$,   E. Mason$^{11}$, 
L.   Girardi$^{1}$, V.      D'Orazi$^{1}$,  G. Piotto$^{2}$,  
A.P.  Milone$^{2}$,  H. Lala$^{2}$,  P.B. Stetson$^{12}$     \& 
Y. Beletsky$^{13}$}

\begin{document}
\maketitle

\begin{affiliations}
\item INAF - Osservatorio    Astronomico    di    Padova,    Vic.  dell'Osservatorio        5,       35122        Padova,       Italy
\item Dipartimento   di  Fisica   e  Astronomia,   Univ.    di    Padova,   V.  dell'Osservatorio 3, 35122 Padova 
\item Instituto de Astrofisica de Canarias, E-38205 La Laguna,  Tenerife, Spain
\item Departamento de Astrofisica, Universidad de La Laguna, E-38206 La Laguna, Tenerife, Spain 
\item European Southern Observatory, Karl Schwarzschild Strasse 2, D-85748 Garching, Germany
\item INAF - Osservatorio Astronomico d'Abruzzo, Via M. Maggini, I-64100 Teramo, Italy 
\item INFN - Sezione di Pisa, Largo Pontecorvo 3, 56127 Pisa, Italy
\item Instituto de Astronom\'ia, Universidad Cat\'olica del Norte, Av. Angamos 0610, Antofagasta, Chile
\item European Southern  Observatory,   Alonso  de   Cordova  3107,   Santiago,  Chile
\item Departamento   de  Ciencias   Fisicas,  Universidad   Andres  Bello,  Fernandez  Concha  700,  Las  Condes,  Santiago,  Chile
\item INAF - Osservatorio Astronomico  di Trieste,  Via G.B. Tiepolo, 11, I-34143, Trieste, Italy 
\item Herzberg  Astronomy and  Astrophysics,  National Research Council, 5071 West  Saanich Road, Victoria, BC V9E  2E7, Canada
\item Las Campanas Observatory, Carnegie Institution of Washington, Colina el Pino, Casilla 601, La Serena, Chile
\end{affiliations}


\begin{abstract}
  Six    decades    and     counting,    the    formation    of    hot
  $\sim20,000-30,000$\,K  Extreme  Horizontal  Branch (EHB)  stars  in
  Galactic Globular Clusters remains one of the most elusive quests in
  stellar  evolutionary theory.   Here we  report on  two
  discoveries shattering their currently alleged stable luminosity.
  The  first EHB  variability is  periodic and  cannot be  ascribed to
  binary evolution  nor pulsation.  Instead,  we here attribute  it to
  the presence of magnetic spots: superficial chemical inhomogeneities
  whose projected rotation induces the variability.
The second EHB  variability is aperiodic and manifests
  itself on  time-scales of years.   In two cases, the  six-year light
  curves display  superflare events a mammoth several  million times more
 energetic than solar analogs.
 We  advocate a  scenario where  the two  spectacular EHB variability
 phenomena are different  manifestations of diffuse, dynamo-generated,
 {\em weak} magnetic fields.
 Ubiquitous magnetic fields, therefore,  force an admittance into the
  intricate  matrix governing the  formation of all EHBs, and traverse
 to their Galactic field counterparts.
 The bigger picture is one where our conclusions bridge similar
  variability/magnetism  phenomena in  {\em all} radiative-enveloped
  stars: young main-sequence  stars, old EHBs and  defunct white dwarfs.
\end{abstract}

\bigskip

We  report  on the  results  of a  monitoring survey  of the  hot
stellar  populations in  three Globular  Clusters (GCs).   The novelty
resides in  exploiting the near-ultraviolet filters  that suppress the
contribution  of  the  (undesired)  bright/cool red  giant  stars  and
enhance that of the faint/hot stars -- the targets of the survey.
These  are  the  $\sim0.48$\,M$_{\odot}$ stars  spanning  temperatures
between  $20,000-30,000$\,K,  referred\cite{brown-discont-2016} to  as
Extreme  Horizontal Branch  (EHB) stars  when identified  in GCs  (and
sub-dwarf B-type, sdB, when found in the Galactic field).
EHBs are believed  to have lost $\sim30\%$ of  their original mass and
almost  the entirety  of  their envelope\cite{castellani1993}.   Their
Hydrogen shell  is so thin ($\lesssim0.02$\,M$_{\odot}$)  that, in the
wake of exhausting their Helium-burning  core,  they are doomed to
  skip  the  Asymptotic  Giant  Branch  (AGB)  phase;  instead  they become
  brighter  and  then  head  directly to  the  white  dwarfs  (WDs)
graveyard.

The vast majority ($\sim80\%$) of the Milky Way sdBs resides in binary
systems;     over     $50\%$      of     which     are     in     close binaries\cite{heber2016}.  
The  unavoidable  interactions  among the  close companions  therefore
devise a viable mass-loss mechanism capable of stripping  the stellar core
of its envelope and forming an sdB star.
On the  other hand,  EHBs in GCs  display a surprising dearth  of binary
systems\cite{chris2006,chris2011,moehler2011,latour2018}.
By all means, of the  hundreds of EHBs  monitored to date,  only {\em one}
 EHB binary system is known\cite{chris2015}.
The  shortage of  EHB binaries  is supported  by recent  {\em spectroscopic}
monitoring of GCs' red giants  inferring a very low ($\lesssim5\%$) binary
frequency\cite{sara2015}.
Thereby,  and despite  numerous scenarios  put forward  throughout the
past   $\sim60$   years\cite{catelan2009,gratton2019}  (involving   an
interplay  between  cluster's  age, cluster's  central  concentration,
Helium-enrichment  and mixing,  CNO abundance,  stellar  rotation, and
extreme  mass-loss)   EHBs  in  GCs   lack  general  consensus   on  a
comprehensive formation scenario.  In this framework, our findings are
likely to  provide new impulse in  tackling their properties.

Figure~1  summarises  the first  of  our  findings:  the detection  of
sinusoidal,  single-wave, EHB light modulations   having  intermediate   periodicities
($\sim2-50$\,days) and low amplitudes ($\Delta\,U_{Johnson}\sim0.04-0.22$\,mag).
The periodic EHB variability is detected in three GCs, spanning a wide range in
metallicity,  dynamical  history,   age,  census  of  multiple-stellar
populations, and estimated Helium-enhancement.
The  above conclusion,  coupled  with the  evident  similarity of  the
variability's periods/amplitudes  and shapes (in three  GCs) allows us
 to  predict that these EHB variables  are ubiquitous among
all GCs possessing EHB stars.   This prediction is already proven true
by independent explorations in other clusters (c.f. Methods).
A   closer   examination   of    the   color-magnitude   diagrams   in
Fig.~\ref{f_cmds}  also  shows  that  the EHB  variables  embrace  a
previously identified photometric discontinuity at $\sim22,500$\,K
(Momany-jump\cite{momany2002}),  and  span   a  temperature  range  between
$\sim18,000-28,000$\,K, that is, the very temperature range of EHBs.
The frequency of EHB variables is inferred by normalising their number
to all  EHB stars confined  within the colour/luminosity range  of the
variables themselves (c.f. Extended Data Fig.\,1).
It  is remarkably  similar: $\sim13\pm4\%$,  $\sim15\pm11\%$, and
 $\sim12\pm5\%$ for NGC2808, NGC6752, and NGC5139, respectively.
 Thus, whatever is  the origin of the EHB variability,  it is likely a
 {\em  universal} phenomenon  and it  performs similarly  in GCs  with
 different properties.

 The origin  of the uncovered  EHB variability is unlikely  related to
   radial  pulsation.   Indeed, having  an extremely  thin H-rich
 envelope,  EHBs  are  relatively  too  rigid  to  allow  radial
 pulsation and, if/when present,  this occurs at very high frequencies
 (multi-periodic, milli-mag oscillations).
  In particular, rapid (pressure-mode) oscillations with periods around
 $\sim100-200$\,seconds   have    been   detected   in    both   field
 sdBs\cite{kilkenny1997}   and   GCs'  EHBs\cite{brown2013,randal2016},
 while   slow   (gravity-mode)   oscillations  with   periods   around
 $\sim0.7-2.0$\,hours\cite{green2003} are solely found in field sdBs.
This  framework   is  simply  irreconcilable  with   the  much  longer
$\sim2-50$\,day periods reported here.
Similarly,  the analysis  of  the  EHBs' light  curves  and  complementary
spectroscopic  radial-velocity  monitoring  (of  few of  the  detected
variables)   allow   us   to   definitively  discard   the binary   origin
(c.f.  Methods and Extended Data Fig.\,2 and 3).
This is again not  surprising, since all previous surveys  searching for EHB
binaries      in      GCs      have      revealed      an      overall
scarcity\cite{chris2006,chris2011,moehler2011,latour2018}.
Having excluded pulsation/binarity, we here propose that the origin of
the    EHB   variability,    according   to    convention\cite{samus2017},   is    the
$\alpha^2$\,Canum\,Venaticorum   ($\alpha^2$\,CVn)   phenomenon,
afflicting  {\it   Magnetic  Chemically  Peculiar  B/A-type}  stars
($B_{P}/A_{P}$).
$B_{P}$ stars  display superficial over-abundance of  certain chemical
elements (mostly Silicon and Iron  and rare-earth elements) that leads
to spatial  variations in  opacity and  subsequent formation  of large
spots, kept stable  (up to several decades) by  an underlying magnetic
field\cite{bernhard2015}.
Eventually,  these  spots rotate  with  the  star, hence  causing  the
photometric/spectroscopic variability.
A defining characteristic of these  magnetic spots is that they should
be {\em dark}  in the Far-ultraviolet and {\em  bright} in the optical
window\cite{miku2019}.

The parallels we draw between the young/massive (few hundred million
years, $\sim1.5-7$\,M$_{\odot}$)    $B_{P}$    stars    and   the  old/low-mass
($\sim12$\,Gyr, $\sim0.48$\,M$_{\odot}$) EHBs might seem bold.
However, besides  sharing the  same superficial temperatures,  we also
note  that   $B_{P}$  and   EHBs  are   characterised  by   a  similar
convective-core/radiative-envelope  structure,  a  similar  $\sim10\%$
frequency   among    their   siblings\cite{bagnulo2002},    and   most
importantly, display similar enhancement/depletion patterns of certain
chemical elements.
The  EHB chemical  peculiarities  are generally  attributed to  atomic
diffusion processes;  where heavier elements are  pushed upwards while
lighter ones are brought downwards.
Indeed,       EHB    stars   straddling    the   M-jump    display
  $\sim10-20$\,times   enhancement   in   Iron\cite{brown-IIjump-2017}
  (relative to the cluster's average metallicity)
and           $\sim5-10$\,times depletion in Helium\cite{moehler2011}
(relative to solar value).
Since all EHBs are  by definition chemically peculiar\cite{heber2016},
the sub-group  of EHB variables  conceivably\cite{paunzen2019} display
un-even surface  chemical distributions,  well-suited to  exhibit $\alpha^2$\,CVn 
spot-induced variability.
Further  confirmation  that the  EHB  variability  is consistent  with
rotating magnetic  spots is served  when translating the  inferred EHB
photometric    periods    into    rotation   rates.    The    inferred
$\lesssim10$\,km/s  are  in  excellent  agreement  with  spectroscopic
studies (c.f. Extended Data Fig.\,4).
Figure~\ref{f_two} summarises our  modelling (c.f.   Supplementary
  Information) of the light curve of variable vEHB-12 in NGC2808.
  The favourable location of this EHB variable (at $7.8^{\prime}$ away
  from  the crowded  and noisier  cluster  centre) has  allowed us  to
  exploit   {\em  both}  the   $U/R$  Johnson-filters   light  curves,
  simultaneously.
  In  particular,   one  immediately  notes   that  the  $U_{Johnson}$
  amplitude   is  unambiguously   {\em  larger}   than  that   in  the
  $R_{Johnson}$  filter.  This  is  in  perfect  accordance  with  the
  requisites      of      the      chemical     spots'      wavelength
  dependency\cite{krti2018}.
  The  simultaneous $U_{Johnson}/R_{Johnson}$  light  curves modelling
  converges on  the presence of  a bright ($\sim2,500$\,K  hotter than
  its surroundings) stellar spot covering as much as $\sim25\%$ of the
  EHB surface.
  Extrapolating from  detailed modeling\cite{shavrina2010,glag2007} of
  field  $B_{P}$ stars  (displaying photometric/spectroscopic/magnetic
  single-wave modulations, c.f.   Supplementary Information) the
  sizeable spot  of vEHB-12  likely reflects the  presence of  an {\em
    off-centre} dipolar magnetic field  whose poles are separated only
  by the spot's diameter ($\sim60^{\circ}$).
    Due  to the  averaging process  over a  given hemisphere,  similar
    magnetic field configuration will  expectedly deliver, an overall,
    {\em weak} field strength measurement.

    We  take our modelling  of the  stellar spots  a step  further and
    address the question: {\em why is the $\alpha^2$\,CVn  variability detected in
    only $\sim13\%$ of the EHBs?}
We simulated  over 2  milion light  curves including  a spot  with all
possible  parameters (e.g.   size, location,  and a  given temperature
distribution) and then searched among these modelled curves which ones
had  a variability  signal just  as strong/high  as that  seen in  the
observed EHB variables.
 Although all simulated  light curves include stellar spots,
  only $\sim33\%$ of these  match the observed variability
  signal.
  Interestingly,   assuming  a    realistic  spot   temperature
  distribution, the frequency of  detectable EHB variables goes down to
  $\sim12\%$  (c.f.    Supplementary Information);  very  close  to  the  empirical
  frequency inferred for the three examined GCs.
 This  ultimately  suggests  that  the  majority  of  the  chemically
  peculiar   EHBs   are   lurking   with   small-sized,   low-contrast
  temperature,  and unfavourably  positioned spots:  a true  plague of
  magnetic spots  and yet one only  recognises the tip of  the plagued
  EHB population.

Detailed    modelling\cite{brown-discont-2016,cassisi2013}    of   hot
($\ge6,000$\,K)  horizontal  branch stars  envisages  the presence  of
surface/sub-surface  convective layers  (namely  the first  ionization
Hydrogen and the first  and second ionization Helium convective zones,
HCZ, HeICZ, and HeIICZ).
The three layers not only reach up towards the surface of the star but
can  also,  as a  function  of  specific  temperatures, merge  and/or
disappear.
Focusing our  attention on hot  ($\sim20,000-30,000$\,K) Hydrogen-rich
EHBs, the  envelope of  these stars is  always radiative  with a
close, sub-surface, HeIICZ layer beneath it.
Surprisingly, this  detailed EHB envelope structure  is {\em mirrored}
in   the   models    of   $0.9-7$\,M$_{\odot}$   young   main-sequence
(MS) stars\cite{cantiello2019}.
In   particular,  modelling   of  the   HeIICZ  convective   layer  in
main-sequence stars proves  it being capable of  producing  low-level
  (less  than few  hundred  Gauss)  dynamo-generated magnetic  fields
which, via buoyancy\cite{cantiello2011}, can  easily reach the stellar
surface and form bright magnetic spots\cite{cantiello2019}.
Of  most interest  to  our  proposed EHB/MS  parallelism  is that  the
modelling  of both  EHB\cite{brown-discont-2016,cassisi2013} and
MS\cite{cantiello2019}  stars  predicts  the  HeIICZ to  approach  the
(still radiative) stellar surface  just around the M-jump temperature,
encompassing the EHB variables.
 Furthermore, the two modelings envisage similar thinness for both
  the HeIICZ and the radiative layer above it, all taking place in the
  upper $\sim1\%$ of the stellar envelope (Supplementary Information).
  Granted  the   above  similarities,  we  argue   that  the  HeIICZ's
  capability of generating  magnetic fields in MS stars  can be safely
  extrapolated  to EHB stars  as well.

Obviously, there are uncertainties in modelling these dynamo-generated
magnetic  fields  (e.g.  the  geometry of  the  resulting  spots,  the
strength and lifetimes of the magnetic fields).
 Moreover,  we have  just started  scratching the  surface of  the EHB
 magnetic field properties which, in
 distant   GCs,  is   hardly  detectable   with  currently   available
 instrumentation.
 Nonetheless,  the   stage  seems  set:   there  are  {\em   four  EHB
   observables} whose onset  is temperature ($\sim22,000$\,K) related:
 (i) the  M-jump photometric discontinuity\cite{momany2002};  (ii) the
 Iron-enhancement  discontinuity\cite{brown-IIjump-2017};   (iii)  the
 onset of significant  scatter in Helium-abundance\cite{chris2012} and
 lastly; (iv) the onset of $\alpha^2$\,CVn  variability and appearance of magnetic
 spots/fields.
 There  were  already   suggestions\cite{brown-IIjump-2017}  that  the
 HeIICZ is the   triggering mechanism of the  first two phenomena
 and  likewise,  we here  argue  it  is  likely  to trigger  the  
   third/fourth, too.
   Our discovery  of an  {\it EHB magnetic spot plague}, is  however the
   most revealing: without  the HeIICZ-generated magnetic fields
   the  consequent superficial  spots  cannot  be stabilised  on
   time-scales  of   years,  i.e.   no  magnetic   fields,  no  stable
   spot-induced EHB variability.

   Additional  insight  into  the   turbulent  nature  of  the  HeIICZ
   approaching  the  upper $\sim1\%$  of  the EHB  stellar radius  may  be
   provided  by our  second  detected  variability:  low-amplitude
   ($\Delta\,u_{SDSS}\lesssim0.1$\,mag) luminosity  transitions, occurring on
   time-scales  of hundreds  of days  only  in stars  hotter than  the
   M-jump  (c.f. Fig.~\ref{f_three} and Extended Data Fig.\,5).
   Seven  such  variables  (which  we  denominate  {\it  Padua})  were
   detected in NGC6752 (the only GC for which we collected a long-term
   $\sim6-$year   monitoring)  and   in  two   cases   the  luminosity
   transitions were preceded by mini-bursts.
 As for the periodic EHB  variables, we also preclude binary evolution
 and stellar pulsation  being the origin of  the aperiodic variability.
 Furthermore, the inferred  frequency of the Padua-variables in
 NGC6752 is  $\sim10\%$,  in line with  that of  the periodic
 EHB variables.

 Padua-1 is an excellent  representative of the entire Padua-family as
 it simultaneously  displays a luminosity transition  and a mini-burst
 event. Both are confirmed in the $r-$filter light curve.
  Since we exclude binary evolution  and pulsation, the most immediate
  framework explaining the Padua-1 mini-burst event is sought in
  magnetic superflare events.
 Figure~\ref{f_three}  shows   the  light  curve  of   Padua-1,
 displaying  a full-sampled  mini-burst whose:  sharp rise/exponential
 slow  decay  form, $\Delta\,u_{SDSS}\sim0.05$\,mag  amplitude,  $\sim80$-day
 lifetime, and  $10^{39-40}$\,erg released energy are  consistent with
 that of  a superflare event.  
 Sun-like   flares  occur   when   convection  in   the  vicinity   of
 spots\cite{schaefer2012} distorts the magnetic  field which, in turn,
 pierces  the  photosphere and  connects  the  corona to  the  stellar
 interior.   This  mechanism  releases   large  quantities  of  stored
 magnetic energy.
 In general, flares last up  to few hours. However, evidence exists of
 superflare   events  enduring   $\sim25$\,days\cite{schaefer2000}  in
 F-type stars  (releasing $\sim10^{37}$\,erg) and  others lasting tens
 of   days\cite{schaefer1989}  in  radiative-enveloped   B-type  stars
 (releasing $\sim10^{40}$\,erg).
 The slow EHB rotation velocities  (as inferred here from our periodic
 variables  and Kepler-based  sdB studies\cite{reed2014})  further provide 
 the necessary grounds for witnessing a long superflare event as
 in Padua-1.
 Furthermore,  the existence  of  such energetic/enduring  superflares
 would not be surprising\cite{balona2015}  especially if the EHB stars
 display large-scale spots coverage (e.g.  Fig.~\ref{f_two}).

 To comprehend the energetics involved in the Padua-1 superflare
 event, one should bear in mind that (even from a Kepler point of view)
 the    spectacular   solar    flare    events    would   still    not
 qualify\cite{balona2015} our  Sun as  a flaring  star (nor  a spotted
 star for that matter).
 And  yet   the  detected  superflare   event  in  Padua-1  is
 $\sim10$\,milion times more energetic. Thus,  the origin of this rare
 and possibly  devastating event is  even harder to  model, especially
 considering its occurrence in radiative-enveloped stars.
Nonetheless,  we stress  the  fact  that energetic  superflares  as  in  
   Padua-1  may  also  configure  a  mass-loss/stellar-wind  process,
 possibly   relevant\cite{fontaine2006}  to   the  low   frequency  of
 pressure-mode oscillators among sdBs 
 and   maybe    to   the    occurrence   of    secondary   helium-core
 flashes\cite{miller2020}.

 In   conclusion,  the   detection  of   the  periodic/aperiodic   EHB
 variabilities  in  GCs has  come  as  a  total surprise.   
 The two  types of  EHB variability are  incorporated within  a single
 theoretical framework and are  substantiated by its predictions: {\em
   weak}  magnetic fields,  of  only few  hundreds  of Gauss,  {\em do} trigger
 discernible observables such as magnetic spots and superflare events.
 In this regard, we emphasise  that the ubiquitous nature of magnetism
 among  EHBs in  GCs is  very much  likely applicable  to sdBs  in the
 Galactic field as well.
In particular, we  argue that  the  distinctive property  of EHBs/sdBs
 (i.e.  dearth ${vs}$ excess in binary frequency) is likely behind the
 concealing of the magnetic spots/fields signature in sdBs.
 To  test  this  hypothesis  we  searched  among  the  small  minority
 ($\sim20\%$)   of   apparently   single-sdB  stars   inspecting   for
 spot-induced rotational-modulations and  possibly flaring events, and
 we   did  find   one  such   trojan-horse:  CD-38\,222   (c.f.   
   Extended Data Fig.\,6).
  Not only does it display $\sim0.22$\,days  rotational light modulation,
 but it   also  exhibits   a   fortuitous  detection   of  an   energetic
  ($10^{35}$\,erg)  superflare  event.    The  latter  constitutes  an
  ultimate proof  of an underlying  magnetic field, regardless  of its
  strength and/or detectability.
  Interestingly,  the magnetic field  in CD-38\,222  was independently
  investigated\cite{landstreet2012,bagnulo2015}  and  inferred  to  be
  below $\sim400$\,Gauss.

  The above arguments  thereby support a scenario where 
    weak magnetic fields are perfectly capable of triggering
  discernible observables.
 These  localised/weak  magnetic fields  are  likely responsible  for
  the puzzling     chemical    anomalies     (e.g.     horizontal/vertical
  stratification)   and   are   an   excellent  candidate   for   the
  (long-searched)   mass-loss   process   and  rotation-braking,   all
  pertinent to the very formation of EHBs in GCs.
   This also  brings about  the possible  role  of localised/weak
    magnetic fields  (and thereby  implied presence of  stellar spots)
    producing other puzzling anomalies in other (convective-enveloped)
    phases\cite{somers2020}.   These  are  particularly  pertinent  to
    old/low-mass stars (similar to those in GCs) for which a confirmed
    weakened magnetic braking is now established\cite{saders2016}.
     What is sure is that  magnetic fields (although weak and not yet
      detectable)  can  no  long  be  considered  an  exotic  feature
      nor a luxury ingredient reserved only for the closest of stars.

Looking  at  an even a bigger  picture,  the  implications  of  the  EHBs'
 universal magnetism  extend to  {\em all}  B-type radiative-enveloped
 stars (i.e.  young  MSs, intermediate-age sdBs, old  EHBs and defunct
 WDs), all showing apparently puzzling rotational photometric variability.
 In  all cases  however, the  puzzle  is/was soon  solved by  invoking
 different shades of $\alpha^2$\,CVn variability  and weak magnetic fields,
 despite the radiative-enveloped stars at hand.
 The  full-circle is  most evident  in the  field of  young  B-type MS
 stars:  passing   from  ``shocking''  early   reports  of  rotational
 variability            detections            to,            nowadays,
 acceptance-calls\cite{balona2019b}  openly demanding  a  {\it ``major
   revision of stellar physics''} for radiative-enveloped MS stars.
Indeed,  Kepler/TESS studies  now  establish\cite{balona2019a} that  a
large   ($\sim40\%$)  fraction  of   B-type  stars   show  photometric
variability with  periods consistent with the  stars' rotation periods
(i.e. spot-induced variability).
 Thus, reluctance  to accept the  proposed magnetic spots'  plague for
 EHBs is likely  to echo the early stages of $\alpha^2$\,CVn detection in B-type MS
 stars.

 Interestingly,  hot radiative-enveloped  WDs are  no exception,  and
 evidence of  the presence of dark/bright  spot-induced variability is
 mounting rapidly (c.f.  Supplementary Information).
 The  parallelism we  draw between  the EHB/hot-WD  photometric spots'
 properties  is likely  to be  conceded easily,  especially  given the
 imminent EHBs' evolution into WDs.
 On  the  other  hand  however,   one  is  expectedly  reluctant  when
 recognising the quite different magnetic properties of EHBs and
 WDs.
  In  particular, one  is  understandably  intimidated when  comparing
  typically  strong  magnetic fields  detected in  WDs (reaching
  hundreds  of  milions  of  Gauss) with  the  allegedly   weak
  magnetic fields (up to few hundreds Gauss) we propose exist in EHBs.
  Enough to mention spotted-WDs cases like GD\,394 or J1529+2928 (with
  magnetic        fields'       estimated        upper-limits       of
  $\lesssim12,000$\cite{dupuis2000}                                 and
  $\lesssim70,000$\cite{kilic2015}\,Gauss,   respectively)   are   yet
  considered non-magnetic WDs.
  However, the  viability of the suggested  EHB/WD magnetic 
  parallelism  is soon  clear once  the EHB-to-WD  {\it  shrinkage} in
  stellar    radius   is   taken    into   account.
  Indeed,   EHBs/sdBs   (with   $20,000\le\,T_{eff}\le30,000$\,K   and
  $\sim0.45\,R_{\odot}$) are incapacitated  to ascend their programmed
  AGB  phase   and  instead  evolve  to slightly  brighter
    luminosities  (i.e. lower  gravities and  larger  radii) following
    which  they bypass  directly to  the  WDs ($\sim0.01\,R_{\odot}$)
  stage.  This detour  typically implies  a  $\sim45$\,times shrinkage
  factor in stellar radius.
  If so, and  assuming that surface magnetic flux  is conserved during
  the collapse, then even  a modestly weak $\sim400$\,Gauss (i.e.  the
  estimated    upper   limit   for    the  sdB    CD-38\,222 field)   would
  translate\cite{koester1990}     into     an     impressive
  $\sim800,000$\,Gauss field acting upon the surface of a WD.
  Thus, the fact that significant spot-induced variability is reported
  in apparently non-magnetic WDs strengthens our proposal that
  weak EHBs' magnetic fields play an important role in {\em
    bridging} their magnetic/spots properties with their hot-WDs
  counterparts. 
  This  is further  supported  by studies\cite{reding2018}  concluding
  that strong WDs' magnetism is not required to produce spots on their
  surface.

  Traditionally, magnetic  WDs were sought  as the descendents  of the
  magnetic $B_{P}/A_{P}$ main-sequence stars,  i.e.  the WDs' magnetic
  fields are fossil\cite{mestel1966}.  However, there
  is  an  ongoing  debate\cite{ferrario2020}  whether  these  magnetic
  fields  can be  dynamo-generated,  in some  later evolutionary
  phase.
  In  this   regard,  our  conclusions  supporting   the  presence  of
  dynamo-generated  magnetic  fields  are  tightly  connected  to  the
  sub-surface  HeIICZ  and its  very-close  proximity  to the  stellar
  surface.
  In particular, this HeII convective layer does not date back
  to the original formation epoch at the main-sequence and is fruit of
  successive evolution (mergers not excluded).
  Hence,  magnetic EHBs  in GCs  provide a  unique benchmark  where to
  investigate the  role of {\em non-fossil}  magnetic fields occurring
  in  currently-single/very-old/low-mass   stars,  on  the   verge  of
  becoming WDs.

\bigskip
\bigskip
\bigskip
\bigskip



\bigskip
\bigskip

\begin{addendum}
 \item 
   We acknowledge  fruitful discussions  with S. Bagnulo,  A. Bressan,
   A. Bianchini, A. Renzini, and P. Ochner, and thank M. Dima for help
   in producing movies of the stellar spots.
DJ acknowledges  support from the  State Research Agency (AEI)  of the
Spanish Ministry  of Science,  Innovation and Universities  (MCIU) and
the   European  Regional   Development   Fund   (FEDER)  under   grant
AYA2017-83383-P.  DJ  also acknowledges  support under  grant P/308614
financed by  funds transferred from  the Spanish Ministry  of Science,
Innovation and Universities, charged to  the General State Budgets and
with  funds transferred  from the  General Budgets  of the  Autonomous
Community of the Canary Islands  by the Ministry of Economy, Industry,
Trade and Knowledge.

\item[Author  Contributions] Y.M.   and S.Z.   designed the  study and
  coordinated the  activity.  Y.M., S.Z., M.M.,  H.M.J.B., D.J., M.G.,
  I.S.,  L.M.,  C.M.B.,  V.D.  and  H.L.   reduced  and  analysed  the
  data. M.M. and S.Z. developed the spot modelling programme and related
  simulations.
  S.C., L.G., and D.J. provided theoretical modelling.
  G.P., A.P.M., P.B.S.,  Y.B. and E.M. contributed to the assembly of
  the photometric catalogs.
  Y.M.  wrote the paper.  S.Z., D.J., H.M.J.B., I.S.,  S.C., L.G.
  and H.L.  contributed to the discussion and presentation of paper.
  All authors contributed  to the discussion of the  results and commented
  on the manuscript.

\item[Competing Interests] The authors declare that they have no
competing financial interests.

\item[Correspondence] Yazan Al Momany~(email: yazan.almomany@inaf.it)

\end{addendum}

\begin{figure*}
\begin{center}
\includegraphics[width=15cm]{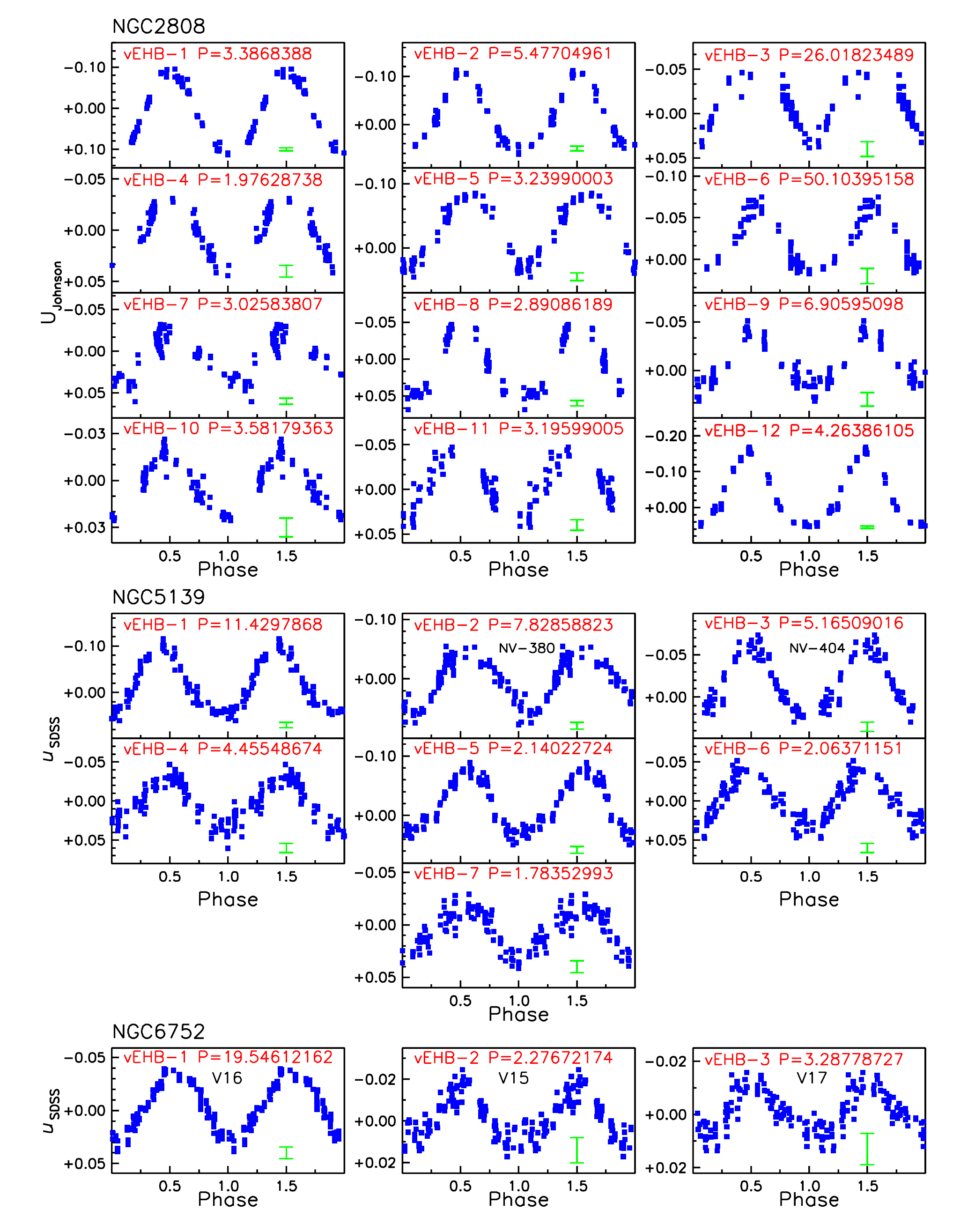}
\caption{The phased near-ultraviolet light curves of the EHB variables
  in three  globular clusters.  The  phased epochs are  repeated twice
  for  a  better  display  of  the  periodicity.   We  used  the  $U_{
    Johnson}$-filter for the NGC2808 data (collected with VIMOS at the
  UT3 telescope) and the $u_{SDSS}$-filter for the NGC6752 and NGC5139
  data  (collected with  OmegaCAM  at the  VST  telescope).  For  each
  variable we report an assigned-identifier, the photometric period in
  days,  and the  typical  $1-\sigma$ photometric  uncertainty of  the
  measurements (plotted as an error bar).}
\label{f_one}
\end{center}
\end{figure*}

\begin{figure*}
\begin{center}
\includegraphics[width=\textwidth]{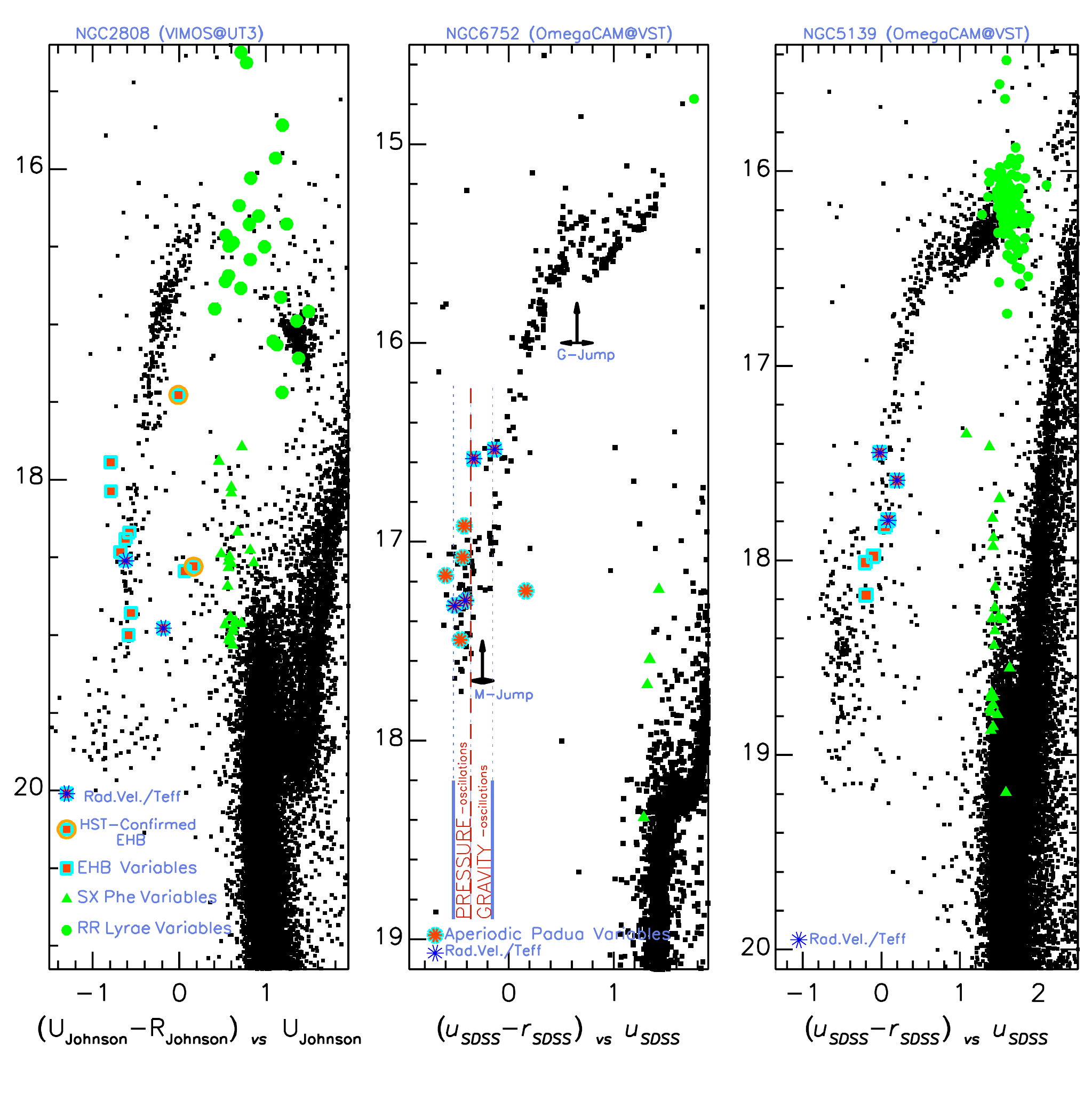} 
\caption{The  EHB variables  in  the near-ultraviolet  color-magnitude
  diagrams.  The periodic EHB  variables are plotted as filled red squares
  while the  long-term aperiodic Padua variables  as filled red asterisks.
Stars with spectroscopic data are further highlighted with blue asterisks.
  Also  plotted are  the known  RR\,Lyrae and  SX\,Phoenicis pulsators
  along   with    the   location   of    the   Grundahl-\cite{grun1999}   and
  Momany-\cite{momany2002}   jumps   (at   $\sim11,500$   and
  $\sim22,500$\,K,  respectively). The vertical  lines in  the NGC6752
  diagram represent the temperature boundaries\cite{pietru2017} of the
  pressure/gravity mode oscillations.}
\label{f_cmds}
\end{center}
\end{figure*}

\begin{figure*}
\begin{center}
\includegraphics[width=0.9\textwidth]{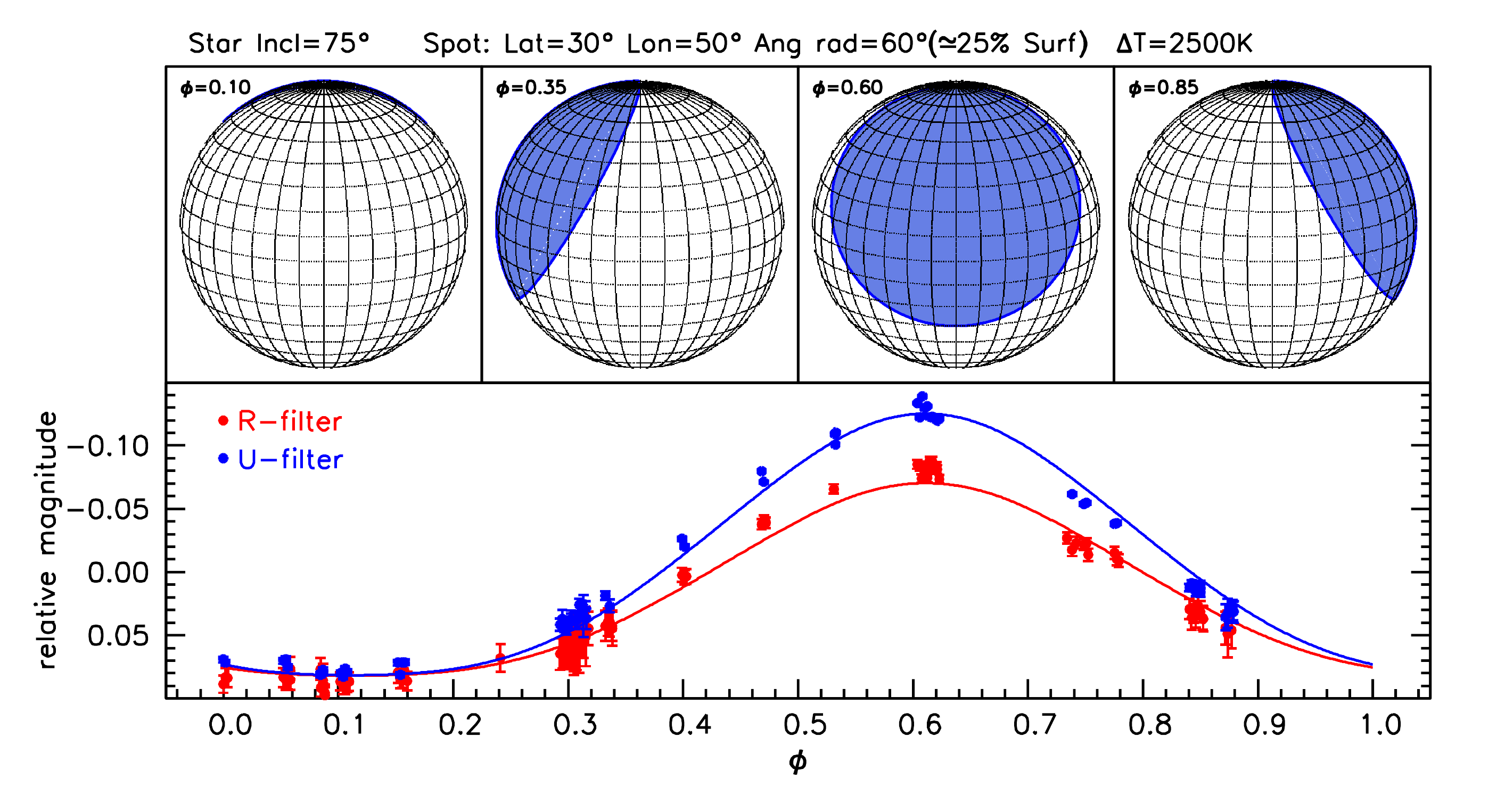}
\caption{Modelling  of the stellar  spot in  the NGC2808  EHB variable
  vEHB-12.   Lower  panel  displays  the  superposition  of  the  $R_{
    Johnson}$ filter (red circles)  and the $U_{Johnson}$ filter (blue
  circles) light curves, along with  the $1-\sigma$ error bars of each
  measurements.   Solid lines show  the corresponding best fitting model
  for  a simulated spotted  star, whose  modelled {\em  bright} spot's
  parameters  (inclination,  longitude/latitude,  angular radius,  and
  temperature  contrast) are  reported.  The  upper  horizontal panels
  display  a snapshot  of the  modelled spot  position  (shaded region
  covering  $\sim25\%$ of  the  stellar surface)  at  four key  phases
  during the $\sim4.3$-day cycle.}
\label{f_two}
\end{center}
\end{figure*}

\begin{figure*}
\begin{center}
\includegraphics[width=\textwidth]{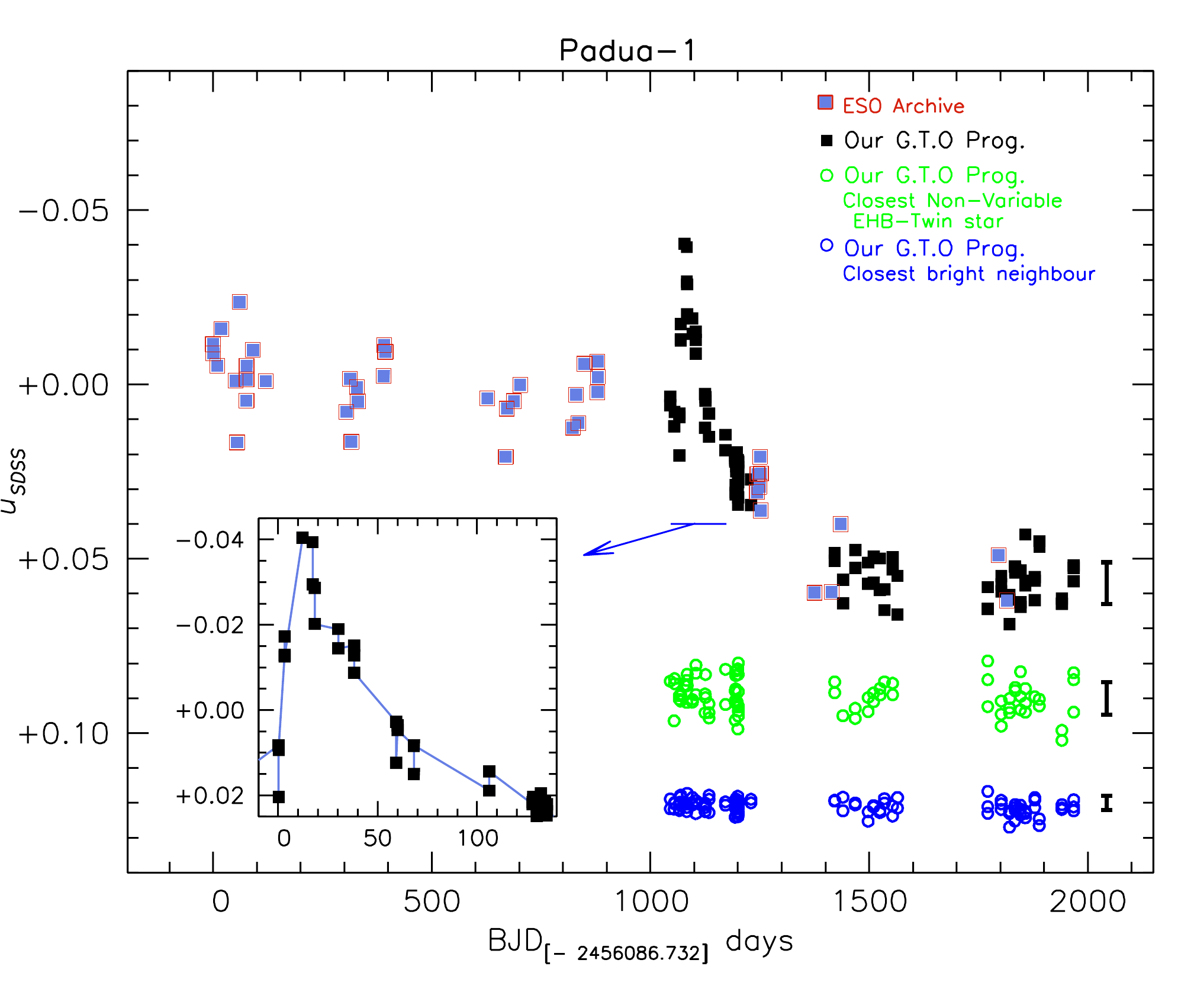} 
\caption{The $u_{SDSS}$ light  curve of  the EHB  Padua-1
  showing a complete superflare event.  Light blue squares display the
  six-year  monitoring of  Padua-1  using archival  OmegaCAM@VST
  data, while  black squares  display our three-year  monitoring using
  the same instrument/filters/exposure  times but collected with
  better  seeing  and more frequent sampling.  
 The agreement between  the two  data-sets is  evident. 
 Two   comparison  stars  (green/blue  open circles) 
 along with their  respective $1-\sigma$ photometric error bars are also plotted.  
 The  constant  trends displayed  by  the  comparison  stars (at  only
 $\sim20^{\prime\prime}$  and  $\sim13^{\prime\prime}$  from  Padua-1)
 prove  that   the  $\sim0.05$\,mag  fainting   and  the  $\sim80$-day
 superflare event are intrinsic to Padua-1.}
\label{f_three}
\end{center}
\end{figure*}

\clearpage


\begin{methods}

\section*{Photometric Data Sets}
\label{s_datasets}
The  presented  light  curves (Fig.~\ref{f_one})  and  color-magnitude
diagrams    (Fig.~\ref{f_cmds})    are    based    on:    (i)    VIMOS
$U_{Johnson}/R_{Johnson}$  photometric  data   collected  at  the  UT3
telescope of the  Very Large Telescope at the  ESO Paranal Observatory
(Chile);  P.I.   D.    Jones  (programme  094.D-0082);  (ii)  OmegaCAM
$u_{SDSS}/r_{SDSS}$ photometric data collected at the VST telescope at
Paranal;   P.I.   Y.    Momany  (programmes   095.D-0307,  096.D-0021,
097.D-0012 and 099.D-0348) and  S.  Zaggia (programmes 0100.D-0023 and
0101.D-0171); and (iii) archival OmegaCAM data (programme 60.A-9038).
{All  data presented were  collected through the  ESO service-mode
  and are available for download from the ESO archive.}
 VIMOS\cite{fevre2003},  de-commissioned as of now, was  a wide-field
  imager made of four identical  arms,  each Quadrant was equipped
  with   a  $4$\,k\,$\times\,2$\,k   CCD   having   a  pixel-size   of
  $0.205^{\prime\prime}$   and    a   usable   field   of    view   of
  $7^{\prime}\times8^{\prime}$ per quadrant.
  We collected  $18$ epochs of  $U_{Johnson}/R_{Johnson}$ exposures of
  NGC2808 spread  between November $20^{th}$ 2014  and March $12^{th}$
  2015.   Overall,  we  collected $68\times120_{sec}$\,images  in  the
  $U_{Johnson}$  filter and  $91\times90_{sec}$  in the  $R_{Johnson}$
  filter,  obtained with  excellent  ($\sim0.7^{\prime\prime}$) seeing
  conditions as measured in the images.
  NGC2808 was  placed in the  center of Quadrant\#1, providing  a full
  coverage  of  the central  $\sim5^{\prime}$  region.  The $3$  other
  quadrants   allowed   us to   sample   NGC2808   populations   out   to
  $\sim18^{\prime}$. A  minimum dithering pattern, consisting  of only
  few arcseconds, was  employed in order to  minimize the introduction
  of artefacts (e.g.  sky concentration, reflection, saturated columns
  etc).
  OmegaCAM\cite{kuijken2011} is the wide-field  imager on the $2.6-$m
  VLT Survey Telescope (VST) telescope.
It  provides   a   total   un-vignetted   field   of   view   of
$1^{\circ}\times\,1^{\circ}$    with   a    32-CCD    mosaic   in    a
$8^{columns}\times4^{rows}$    pattern.      Each    CCD     covers    a
$7.3^{\prime}\times14.6^{\prime}$  field  of  view,  having a  pixel  size  of
$0.21^{\prime\prime}$.
For NGC6752 we collected $104\times300_{sec}$ and $89\times90_{sec}$
$u_{SDSS}/r_{SDSS}$  exposures,  respectively.   These  had  an  average  seeing  of
$\sim1.3^{\prime}$ and $\sim1.0^{\prime}$  covering the period between
April-2015 and  October-2018.  For NGC5139  we analysed $89$  and $85$
$u_{SDSS}/r_{SDSS}$ images collected between April-2015 and May-2017.
The  two clusters  were centered  in  CCD\#84 providing  a catalog  of
$1^{\circ}\times\,1^{\circ}$.
Being  part of  the calibration  plan of  the Paranal  Observatory, we
complement  our   NGC6752  OmegaCAM  data-set  with   archival  images
collected  with the  same  instrument/filters/exposure times,  between
April-2012 and July-2017.
The  homogeneity of  the  archival data-set  allowed  the recovery  of
$98/90$    images   with    seeing    of   $1.4^{\prime\prime}$    and
$1.1^{\prime\prime}$, in $u_{SDSS}/r_{SDSS}$ respectively.  
For all OmegaCAM data we made sure that each scientific image was 
  de-biased  and     flat-fielded  with   proper  images
collected within $\sim2-3$\,days from the observations  themselves.
The entire archival data-set of NGC6752 was reduced independently.
Our monitoring of  the $3$ GCs is still ongoing. We emphasise that
NGC6752 is the only GC for  which we present $u$ monitoring spanning
a period of $\sim6$\,years, and it is the only one where the long-term
variables have been properly identified/investigated.

  The photometric  reduction of the VIMOS/OmegaCAM  dat-sets was based
  on  Point-Spread  Function  (PSF)   photometry  that  was  performed
  independently  on  each  single  image using  the  PSF-fitting  code
  {\sc DAOPHOT\,ii/ALLFRAME}\cite{pbs1987}.
  The   instrumental   photometry    was   calibrated   using   linear
  transformations based on measurements of Peter Stetson complementary
  photometric standard stars.
  Our  final photometric/astrometric  catalogs  of the  three GCs  are
  photometrically  complete  down  to  $\sim2$\,magnitudes  below  the
  cluster's   main  sequence  turnoff   level,  and   artificial  star
  experiment   indicated   a   $\sim50\%$   completeness   levels   at
  $\sim3$\,mag  below  the  turnoff  levels  (c.f.   the  deepness  of
  diagrams in  Fig.\ref{f_cmds} and error  bars in the  light curves).
  This ensured  optimal high-precision photometry  and completeness of
  the EHB and the blue hook stars when present.

  The  light  curves were  constructed  using  the differential  image
  subtraction   (ISIS2.2)  technique\cite{alard1998}   which  is
  optimal for the study of crowded fields.
 Its  main  advantage  is  that  it  does  not  assume  any  specific
  functional shape  for the PSF of  each image. Instead it  models the
  kernel that convolved the PSF of a reference image (created from the
  best  seeing images)  to  match  the PSF  of  a  target image.   The
  reference  image  is  convolved  by the  computed  kernel  and  then
  subtracted from all, single images. We emphasise that the reference
  image was constructed  regardless  of the specific epoch of the
  best-seeing images, hence ensuring a homogeneous treatment of images
  collected over different observing-seasons.
The final  photometry on the  subtracted images was performed  with a,
properly   modified,   {\sc   DAOPHOT}     aperture-photometry
routine\cite{montalto2007}.
The  aperture photometry radius  and inner/outer sky annuli
were  set for  each  single  image separately,  based  on the  average
full-width at half-maximum of the respective image.
The  absolute  time  stamp  reported  in  our  light  curves  are  the
Barycentric               Julian              Date               (BJD\cite{bjd})
which is  referenced to the  center of mass  of the solar  system, and
thereby  corrects for  the  Sun's movement  due  to the  gravitational
attraction of the planets.

The  light  curves  were  examined for  variability  running  the  {\sc   AoV} 
(Analysis of  Variance), {\sc   AoV$_{-}$harm }  (Analysis of Variance  using a
multi-harmonic  model) and  {\sc LS}  (Generalised  Lomb-Scargle search  for
periodic  sinusoidal signals)  algorithms  -- all  found  in the  {\sc
  VARTOOLS}     package\cite{vartool,schwa1,schwa2}.
Thus, for each  light curve we searched for periodic  signals using the
three different algorithms.
The average  power of the entire  sample was derived,  and stars
with  periodic signals  power  exceeding $3$  times  the average  were
folded with their respective period estimates and examined in detail.
In particular, the strongest $5$-peaks (from each of the $3$ algorithm
analysis) were  examined in  order to check  for possible  aliases. In
basically all cases, the peak delivering the least residuals coincided
with  that being  the  strongest and  the corresponding   {\sc   AoV$_{-}$harm }  
period was assumed.
All  reported   variables  were   visually  inspected   for  neighbour
contamination.  In  particular, the  light  curves  of all  neighbours
within $\sim20^{\prime\prime}$  were thoroughly examined  to ascertain
that  the   true  variability  source is  the  one  we  list  in
 Supplementary Table\,1.
 This      process      was       repeated      for      both      the
 $U_{Johnson}/R_{Johnson}$-filters  data independently.   We emphasise
 however that  our variability search/analysis (for all  3 GCs) relies
 primarily on the $U_{Johnson}/u_{SDSS}$-filter light-curves, as these
 allowed  us   to  suppress   the  contribution  of   the  (undesired)
 bright/cool red giant stars and  enhance that of the faint/hot stars,
 the  targets   of  the  survey.    Indeed,  and  especially   in  the
 central/crowded regions, the  $R_{Johnson}/r_{SDSS}$-filter light curves suffered from
 the high-background  and saturation effects which  reflected in light
 curves with lower photometric precision.
Among the EHBs  presented in Fig.~\ref{f_one} there were  few cases of
previously      identified\cite{kaluzny2004,kaluzny2009}     variables
(highlighted  with  their original  identifier  in   Supplementary
  Table\,1  and  Supplementary  Fig.~\,7)   all  of  which  were  {\em
  originally} attributed to binary systems.
  Interestingly, the  estimated periods of the  common variables shows
  an      excellent      agreement      (i.e.       $\sim2.2$      and
  $\gtrsim8$\,days\cite{kaluzny2009}   for   vEHB-2/1  in   NGC6752,   and
  $\sim5.1$  and  $\sim7.1$\,days\cite{kaluzny2004}  for  vEHB-3/2  in
  NGC5139).
  Additional  examples of  intermediate-period, single-wave,  sinusoidal
  EHBs variables are found in NGC6656\cite{rozy2017} ($\sim2.2$\,days)
  and NGC6254\cite{rozy2020} ($\sim0.9$  and $\sim4.5$\,days), in line
  with the inferred  ubiquitous  nature, in all GCs hosting EHB
  stars.

Estimating  the  NGC2808  EHB  variables frequency  is  summarised  in
 Supplementary Fig.\,1. 
In particular, the two EHB variables (vEHB-7
and  vEHB-11  in orange  open  symbols)  were properly  identified  in
higher-resolution       Hubble       Space       Telescope       (HST)
catalogs\cite{milone2015}  and allow  us to  confidently include  them
(and the 2  variables they enclose) within the  final NGC2808 variable
sample.
  To infer the  EHB variable frequency we define  the EHB normalising
  sample as  that delimited by  the colour/magnitude of  the variables
  themselves.
  Indeed, as  shown in  Supplementary Fig.\,1, the  EHB distribution
  shows     the      clear     presence     of      two     previously
  studied\cite{ferraro1998,rolly2000}        EHB        gaps        at
  $R_{Johnson}\simeq18.5$    and   $\simeq20.0$,    corresponding   to
  $\sim17,000$\,K and $\sim25,000$\,K, respectively.
The  occurrence/temperature of these gaps is  subject to change based on
  the  cluster being  examined and/or  the filter-set  being employed.
  Hence, the  safest selection of  the EHB normalising sample  is that
  designed by the EHB variables.
  Overall,   the  NGC2808   (periodic)  EHB   variable   frequency  is
  $\sim13\pm4\%$ ($12/94$), and repeating the same process for NGC6752
  and  NGC5139 we obtain  $\sim15\pm11\%$ ($2/13$)  and $\sim12\pm5\%$
  ($7/60$),  respectively.  These  EHB frequencies  may be  subject to
  small  fluctuations; due  to the  inclusion of  additional potential
  variables  (see  Supplementary Table\,1)  or  the  selection of  the
  normalising box colour/luminosity limits.
  Yet, overall, the  EHB variable frequency for the  three GCs is very
  similar, within $\sim12-15\%$.
  NGC6752 is a particular case  where the two identified/confirmed EHB
  periodic   variables   span   a    very   limited   range   in
  luminosity. Interestingly however, the corresponding normalising EHB
  population  ($\sim13$ stars)  would  still deliver  an EHB  variable
  frequency  of   $\sim15\%$.   If  we   were  to  include   both  the
  periodic/aperiodic EHB variables (i.e.  assume that both variability
  phenomenon represent the  stable/eruptive manifestations of magnetic
  field)   the   resulting   NGC6752  EHB   variables   frequency   is
  $\sim11\pm4\%$ [$(2+7)/81$].

  The temperature range of the EHB variables is inferred from the
    identification   of   few   of  our   variables   in   independent
    spectroscopic studies. In particular, 
 for NGC2808  vEHB-1 and vEHB-5  we  infer $\sim20,500$\,K
  (see  next  Section)  and  $\sim17,900$\,K  (holding\cite{chris2011}
  ID$=42482$),       respectively.       In       NGC5139,      vEHB-2
  (holding\cite{moehler2011} ID$=86429$) has $28,200$\,K, while vEHB-7
  and  vEHB-4  (holding\cite{latour2018} ID$=5255164$  and  $5275033$)
  have $23,829$\,K and $24,494$\,K,  respectively. 
  For  Padua-1 in NGC6752,  we obtain (see   Supplementary
    Information)  $29,800$\,K. 
  Overall, the  EHB variables  (of all three  GCs) span  a temperature
  range between  $\sim18,000-30,000$\,K; i.e.  closely  resembling the
  very definition  of EHBs' temperature range, and  never reaching the
  blue hook temperatures.
  In  this regard,  the theoretical  ZAHB\cite{pietrinferni2013} model
  (used  to  constrain  the  temperatures  of the  two  gaps  in  
    Supplementary  Fig.\,1)  further  confirms the  above  inferred
  temperature range of the EHB variables in NGC2808.
 Lastly, the  case of vEHB-5 in NGC2808  and vEHB-1/2 in NGC6752
    caution that the  variability may extend to the  hotter end of the
    blue HB;  i.e.  slightly  cooler than the  $\sim20,000$\,K nominal
    EHB cool-boundary.

    The lower  panels of Supplementary Fig.\,1 show  the position of
    all the EHB variables we  could identify in available HST catalogs
    for   NGC2808\cite{milone2015},  NGC6752\cite{nardiello2018},  and
    NGC5139\cite{cool2013}.
    Thanks  to  their higher-resolution,  the  HST  diagrams show  the
    disappearance of  the horizontal  bridge connecting  the EHB to
    the MS  turn-off level (likely  caused by  photometric blends
    with   undetected   faint   companions   or   superposition   with
    foreground/background stars).
    In this regard, the HST  diagrams un-ambiguously show that the EHB
    variables are  anchored to  the  mean  EHB locus;  no hints  of any
    particular red-excess.
    Lastly, we note  that if one is to,  inappropriately, derive
    EHB variables  frequency using  the HST  diagrams one  would still
    infer frequencies consistent with  those based on our ground-based
    survey: $\sim16\pm8\%$ and $\sim10\pm6\%$ for NGC2808 and NGC5139,
    respectively.

\section*{EHB Variables Are Not Binary Systems}
\label{s_vrad1}
 For a few of our EHB variables we present specific evidence supporting
  the   incompatibility  of   binary  evolution   with  the   detected
  photometric variability.
 Variable  vEHB-1 in  NGC2808 is  perhaps the  most representative
  case; since  we derived  its spectroscopic parameters  and monitored
  its radial velocity over a period matching its photometric period.
 In particular, a  DDT (293.D-5013, P.I.  Jones,  D.)  of
$1$  hour  was  granted  to  obtain optical  spectra  with  the  FORS2
spectrograph mounted on the UT1 telescope.
Operating through the multi-slit (MXU) mode, on June $23^{rd}$ 2014 we
collected $2\times1350_{sec}$  exposures using the  $GRIS-600B$ grism
(operating between $3300-6210$\,\AA) with a slit width of
$1^{\prime\prime}$ and a resolution of $5.9$\,\AA.
The observations were carried out under excellent seeing conditions of
$0.8^{\prime\prime}$, and  the scientific  frames were  de-biased, flat-fielded,
and wavelength-calibrated using the FORS2 Reflex-pipeline.
The spectra  were then  extracted
with standard  IRAF2 routines,  corrected for  the sky  background and
flux-calibrated.   To  derive   the  effective  temperatures,  surface
gravities, and helium abundances, the observed Balmer and Helium lines
were  fitted\cite{chris2007}  with  properly  selected  stellar  model
atmospheres.
The radial velocity of vEHB-1 confirms  it is an NGC2808 member, and we
obtain    $T_{eff}=20,500\pm\,2,000$\,K,     $logg=5.2\pm\,0.3$    and
$log$(He/H)$=-1.7\,\pm\,0.3$.   The relatively  high  gravity and  low
Helium abundance, confirm that the vEHB-1 is a normal EHB star.

As for the radial velocity monitoring, 
a total of  nine spectra were collected  (programme 094.D-0363) during
three succesive  nights between the $11^{th}$  and $13^{th}$ December,
2014,  using the  FLAMES  facility  mounted at  the  UT2 telescope  in
Paranal.
 The programme,  originally\cite{dorazi2015} aimed at
  inferring Li  and Al  abundances of  red giants  in NGC2808  used the
  HR15N  high-resolution  (R$=\,17000$)  setup,  which  simultaneously
  covers the  H$_{\alpha}$ and  the Li doublet  at $6708$\,\AA.
  Besides  vEHB-1, and  for comparison  reasons, we  monitored  a
  photometrically non-variable EHB, and a known RR\,Lyrae variable.
  A radial velocity  synthetic template\cite{coelho2014} was carefully
  selected and  the standard  cross-correlation technique was  used to
  measure the  radial velocity of  each of  the $9$ single  spectra of
  each star.
 The error of  the single cross-correlation fit does  not reflect the
  actual error  on the radial  velocity measurement, and  therefore we
  associate the error on each measurement  as that relative to the RMS
  of  the $9$  radial velocities,  for each  star.  
  For the programme stars ($V_{Johnson}\simeq16.5$)  the RMS of the heliocentric
  velocity  is $\sim0.8$\,km/s,  and  reached $\sim3.5$  km/s for  our
  relatively  fainter  ($V_{Johnson}\simeq19$) EHB  stars, reflecting  the
  natural increase in the RMS as a function of the decreasing S/N. 
  The  upper panel  of  Supplementary Fig.\,2  proves the  succesful
  detection of  velocity variations  in the comparison  RR\,Lyrae star
  and  shows   an  excellent  agreement  between   the  newly  derived
  spectroscopic    period    and    its    photometric    period    of
  $\sim0.589$\,days.
The lower  panel however,  shows that  the RMS  of the  measured radial
velocities of  vEHB-1 ($\sim2.5$\,km/s)  is comparable to  the typical
$\sim3.5$\,km/s errors, estimated for single measurements.

Similarly,  Supplementary Fig.\,3  presents similar  radial velocity
monitoring  collected for the  EHB variables  in NGC6752.  Relative to
those in  NGC2808, the NGC6752 EHB variables  are $\sim2.5$ magnitudes
brighter ($V_{Johnson}\sim16.5$).
  The  spectroscopic   data  consist   of  $6$   FLAMES  archival
  $2775_{sec}$ spectra (programme  099.D-0527) collected between June,
  $22^{nd}$ and June $30^{th}$, 2017.
  The NGC6752  EHB photometric variables  and the original  targets of
  the  programme  (main  sequence  stars)  basically  share  the  same
  luminosity, hence, the S/N of both types of stars are similar.
 The spectra were collected through the {\it blue} LR02 setup
  ($R\sim6000$) operating between $3960-4570$\,\AA.  This spectral
  coverage thus included the $H_{\gamma}$ and  $H_{\delta}$ and
three Helium lines.
  We therefore  employed a  minimisation method, comparing  the target
  spectrum with a library of synthetic spectra\cite{coelho2014}.
Once again we use the RMS distribution as a tracer of error associated
to each of the $6$ measurements, which for our EHB variables is
around $\sim3$\,km/s. 
The comparison  SX\,Phoenicis   photometric variable  shows clear
signs of velocity modulations whereas the vEHB-1/2 variables { (and
  the potential  EHB candidate  vEHB-4/V17)}, again,  display velocity
RMS comparable to the error estimated for single measurements.

Lastly, no  significant radial velocity variations  were independently
inferred  for  two  EHB  variables included  in  published  monitoring
surveys:  (i) vEHB-5  in NGC2808\cite{chris2011};  and (ii)  vEHB-2 in
NGC5139\cite{moehler2011}.
Thus, it  is rather unlikely  that the  EHB variability is  related to
close-binary evolution.
 One may  argue that  within the achieved  ($\lesssim4$\,km/s) radial
  velocity precision there still  may hide the  imprint of
  low-mass  stellar/sub-stellar  companions  that  introduce  the  EHB
  variability.
  Indeed, the general shape of  the EHB light curves
  seem fairly typical for an  irradiated binary, i.e.  a binary system
  where a hot star irradiates the near face of a cool companion (e.g.,
  the case of HW\,Vir binaries).  Nonetheless, one is readily reminded
  that  the $\sim3.3$-day  period of  vEHB-1@NGC2808 is  $\sim4$ times
  longer  than the  longest  ($\sim0.75$\,days\cite{heber2016}) period
  member of this class.

  Examined  in further  detail, in  an irradiated  binary, it  is the
  differing  projection   of  this  heated  face   that  produces  the
  sinusoidal shape of the light curve.
The amplitude  of an observed irradiation effect  is heavily dependent
on multiple  factors including the  observed wavelength, temperature
difference  between the  two  components, orbital  separation, and  the
stellar radii\cite{jones2017}.
Our  analysis of  the  spectra of  vEHB-1@NGC2808  indicates that  its
temperature  is  of  the order  $\sim20,500$\,K. It  is safe to conclude
that  it  should be  the irradiating  star in  this hypothetical
irradiated  binary.  In  order to  test this  hypothesis, an  array of
models   were   produced   using   the   binary   modelling   software
PHOEBE2\cite{prsa2016} each  of which  comprised a  primary consistent
with  the observed  parameters of  vEHB-1@NGC2808 in  a $\sim3.3$-day
orbit with a  MS companion of varying spectral  type (whose parameters
have  been adopted  from  a  set of  zero  age  main sequence  stellar
models\cite{bertelli2008}).
It  is clear  from  the  observed light curves  that  no eclipses  are
present,  so  orbital  inclinations  were  limited  to  be  less  than
$75^{\circ}$ and  any models which presented  eclipses were discarded.
In all  cases, the amplitude  of the simulated irradiation  effect was
much  lower  than that  observed  for  vEHB-1@NGC2808: for  example  an
approximate K0  companion at  an  inclination  of $75^\circ$  displays  a
semi-amplitude of approximately $\sim0.02$ mag in the $R-$band.
This raises  the possibility  that rather than  being a  main sequence
star, the companion  could be a red giant.  However,  if the companion
would be  an RGB star, it  would almost  certainly be more  luminous than
vEHB-1,  a  possibility  that  is   ruled  out  by  the  robust
identification of the star in the WFC3@HST catalog and its position in
the ground/space-based diagrams.
Furthermore, in all of the aforementioned models, the amplitude of the
effect was,  as expected in  the irradiation scenario, greater  in the
$R-$band  than in  the  $U-$band\cite{hillwig2016}, at  odds with  the
observed  $U_{Johnson}/R_{Johnson}$ light  curves amplitude  (see also
Fig.~\ref{f_two}).
This evidence  allows us to discard  the possibility that vEHB-1  is an
irradiated binary (irrespective of whether the secondary is a low-mass
MS-star or even a planet).
Lastly, concerning the  possible detection of planets  around EHBs (as
is now frequently  observed around white dwarfs) we  note that several
ground/space based surveys aimed at determining the occurrence rate of
planets      in       GCs      have      yielded             null
detections\cite{gilli2000,valerio2012,wallace2020}. 
This  dearth  of  planets  has  been  attributed  to  two  distinctive
properties  of  GCs;  their   low-metallicity  and  the  high  stellar
densities.

An alternative binary  interpretation for the light curve  is that the
variability actually  arises from a so-called  ellipsoidal modulation,
whereby one component  is so close to filling its  Roche-lobe that
the varying  projection of  its tidally distorted  shape reflects an
episodic modulation  in the  light curve. 
In this scenario, the  Roche-lobe-filling star must be  the EHB star -
otherwise the  modulated companion would  be dominant in  the observed
spectroscopy  - and  the orbital  period  must be  twice the  apparent
photometric period (as ellipsoidal  modulation produces two minima per
period\cite{santander2015}).
This would offer a natural  explanation for the almost similar $R_{Johnson}$ and
$U_{Johnson}$ amplitudes as the tidal distortion does not alter with the adopted
photometric  passband.    However,  this  interpretation   is  clearly
unfeasible given that the mass  of the companion required to introduce
significant tidal distortion in the EHB for a period of $\sim6.8$ days
is   prohibitively  large,  while   its  radius   would  have   to  be
exceptionally  small in  order to  not  overflow its  own Roche  lobe.
Therefore, we  can also discard ellipsoidal modulation  as a potential
source of the EHB photometric variability.

Thus, analysis  of the period  and $U_{Johnson}/R_{Johnson}$ amplitude
of  vEHB-1@NGC2808  seems to  preclude  the  viability  of either  the
irradiated/ellipsoidal  binary  scenario  being  responsible  for  the
observed luminosity variations.
This  analysis,  coupled  with  the lack  of  any  significant  radial
velocity variations, allows us to firmly preclude binarity.
Obviously,  eventual  presence  of  companions  in  wide-orbits  (e.g.
orbital period  of $\sim1000$\,days\cite{vos2018}) cannot  be excluded
at the moment.  However, besides  the fact that such wide-binaries are
easily destroyed in the  dense GC environment, their eventual presence
is unlikely to be the cause of the strong and $\sim3$-day short-period
modulations detected here.

\end{methods}


\bigskip
\bigskip
\bigskip

\begin{addendum}

\item[Data   Availability]  All   the   raw  data   (and   associated
  calibrations) used in  this paper are available for  download in the
  ESO Science archive under the respective programme ID (see Methods),
  at \\ \texttt{http://archive.eso.org}. Processed data supporting the findings of
  this study are available from the corresponding author upon request.

\item[Code   Availability] All the codes used in this study are available:\\
Phoebe: \texttt{http://phoebe-project.org/} \\
KSint: \texttt{http://eduscisoft.com/KSINT/index.php}\\ 
EXOFAST: \texttt{http://astroutils.astronomy.ohio-state.edu/exofast/limbdark.shtml} \\
ISIS: \texttt{http://www2.iap.fr/users/alard/package.html}\\
VARTOOL: \texttt{https://www.astro.princeton.edu/$\sim$jhartman/vartools.html}\\
SM: \texttt{https://www.astro.princeton.edu/$\sim$rhl/sm/}\\
ALAMBIC: \texttt{https://esosoft.univie.ac.at/software/esomvm/}\\
DAOPHOT: \texttt{http://www.star.bris.ac.uk/$\sim$mbt/daophot/}

\end{addendum}


\bigskip
\bigskip
\bigskip

%
%
%

\clearpage

\includepdf[pages={1-22}]{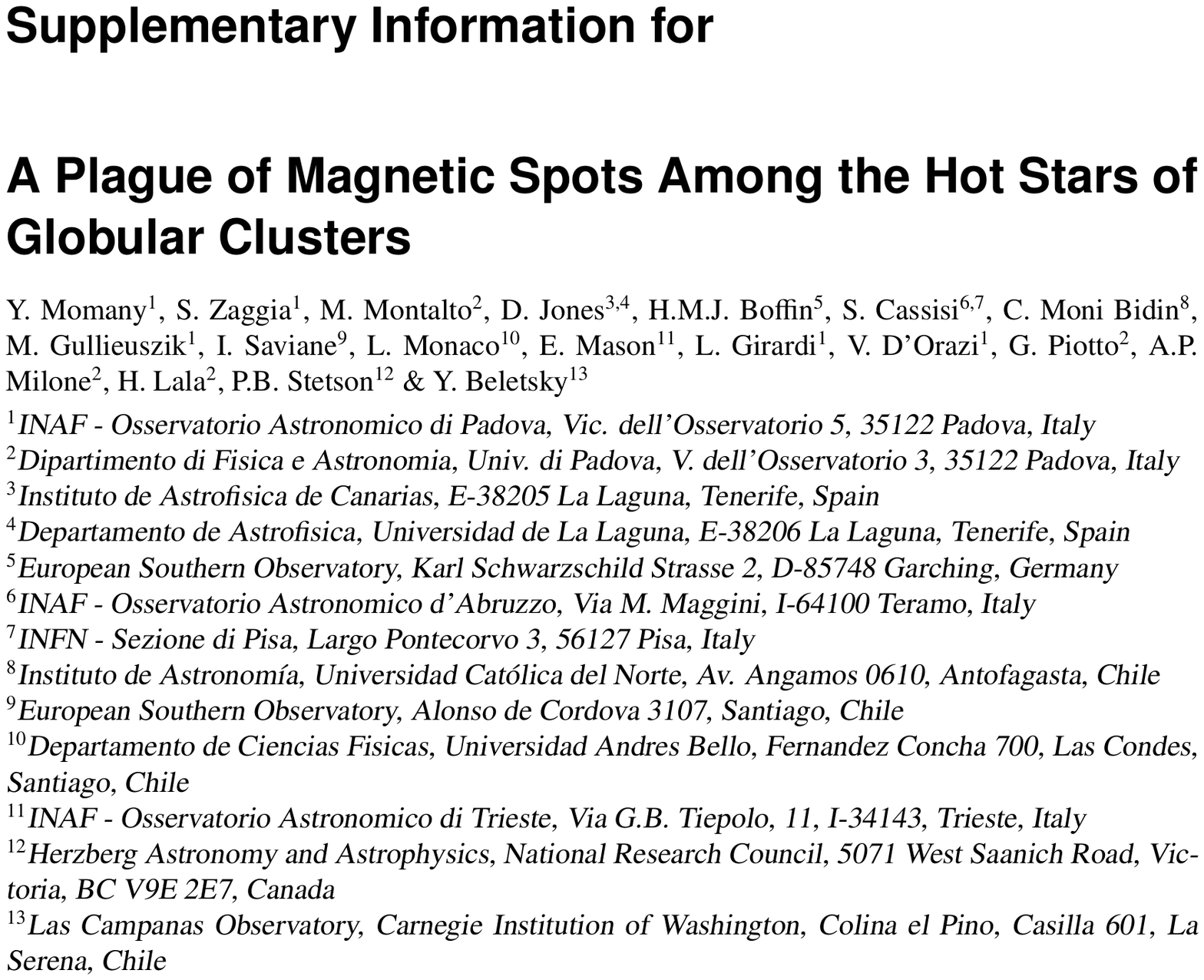}


\begin{thebibliography}{1}
\bibitem{brown-discont-2016}  Brown, T.~M., et al.\ The Hubble Space Telescope UV Legacy Survey of Galactic Globular Clusters. VII. Implications from the Nearly Universal Nature of Horizontal Branch Discontinuities.\, \apj {\bf 822}, 44 (2016).
\bibitem{castellani1993}      Castellani, M., \& Castellani, V.\ Mass Loss in Globular Cluster Red Giants: an Evolutionary Investigation.\, \apj {\bf 407}, 649 (1993).
\bibitem{heber2016}           Heber, U.\ Hot Subluminous Stars.\, \pasp {\bf 128}, 082001 (2016).
\bibitem{chris2006}           Moni Bidin, C., et al.\ The lack of close binaries among hot horizontal branch stars in NGC 6752.\, \aap {\bf 451}, 499-513 (2006).
\bibitem{chris2011}           Moni Bidin, C., Villanova, S., Piotto, G., \& Momany, Y.\ A lack of close binaries among hot horizontal branch stars in globular clusters. II. NGC 2808.\, \aap {\bf 528}, A127 (2011).
\bibitem{moehler2011}         Moehler, S., et al.\ The hot horizontal-branch stars in {\ensuremath{\omega}} Centauri.\, \aap {\bf 526}, A136 (2011).
\bibitem{latour2018}          Latour, M., Randall, S.~K., Calamida, A., Geier, S., \& Moehler, S.\ SHOTGLAS. I. The ultimate spectroscopic census of extreme horizontal branch stars in {\ensuremath{\omega}} Centauri.\, \aap {\bf 618}, A15 (2018).
\bibitem{chris2015}           Moni Bidin, C., et al.\ A Hot Horizontal Branch Star with a Close K-type Main-sequence Companion.\, \apjl {\bf 812}, L31 (2015).
\bibitem{sara2015}            Lucatello, S., et al.\ The incidence of binaries in globular cluster stellar populations.\, \aap {\bf 584}, A52 (2015).
\bibitem{catelan2009}         Catelan, M.\ Horizontal branch stars: the interplay between observations and theory, and insights into the formation of the Galaxy.\, \apss {\bf 320}, 261-309 (2009).
\bibitem{gratton2019}         Gratton, R., et al.\ What is a globular cluster? An observational perspective.\, \aapr {\bf 27}, 8 (2019).
\bibitem{momany2002}          Momany, Y., et al.\ A New Feature along the Extended Blue Horizontal Branch of NGC 6752.\, \apjl {\bf 576}, L65-L68 (2002).
\bibitem{kilkenny1997}        Kilkenny, D., Koen, C., O'Donoghue, D., \& Stobie, R.~S.\ A new class of rapidly pulsating star - I. EC 14026-2647, the class prototype.\, \mnras {\bf 285}, 640-644 (1997).
\bibitem{brown2013}           Brown, T.~M., Landsman, W.~B., Randall, S.~K., Sweigart, A.~V., \& Lanz, T.\ The Discovery of Pulsating Hot Subdwarfs in NGC 2808.\, \apjl {\bf 777}, L22 (2013).
\bibitem{randal2016}          Randall, S.~K., et al.\ Pulsating hot O subdwarfs in {\ensuremath{\omega}} Centauri: mapping a unique instability strip on the extreme horizontal branch.\, \aap {\bf 589}, A1 (2016).
\bibitem{green2003}           Green, E.~M., et al.\ Discovery of A New Class of Pulsating Stars: Gravity-Mode Pulsators among Subdwarf B Stars.\, \apjl {\bf 583}, L31-L34 (2003).
\bibitem{samus2017}           Samus', N.~N., Kazarovets, E.~V., Durlevich, O.~V., Kireeva, N.~N., \& Pastukhova, E.~N.\ General catalogue of variable stars: Version GCVS 5.1.\, \astrep {\bf 61}, 80-88 (2017).
\bibitem{bernhard2015}        Bernhard, K., H{\"u}mmerich, S., Otero, S., \& Paunzen, E.\ A search for photometric variability in magnetic chemically peculiar stars using ASAS-3 data.\, \aap {\bf 581}, A138 (2015).
\bibitem{miku2019}            Mikul{\'a}{\v{s}}ek, Z., et al.\ An Overview of the Properties of a Sample of Newly-Identified Magnetic Chemically Peculiar Stars in the Kepler Field.\ In Kudryavtsev, D.O., et al. (eds.) {\it Physics of Magnetic Stars}, vol. 518 of {\it ASP Conference Series}, 117-124 (2019).
\bibitem{bagnulo2002}         Bagnulo, S., Landi Degl'Innocenti, M., Landolfi, M., \& Mathys, G.\ A statistical analysis of the magnetic structure of CP stars.\, \aap {\bf 394}, 1023-1037 (2002).
\bibitem{brown-IIjump-2017}   Brown, T.~M., et al.\ A Universal Transition in Atmospheric Diffusion for Hot Subdwarfs Near 18,000 K.\, \apj {\bf 851}, 118 (2017).
\bibitem{paunzen2019}         Paunzen, E., et al.\ Search for stellar spots in field blue horizontal-branch stars.\, \aap {\bf 622}, A77 (2019).
\bibitem{krti2018}            Krti{\v{c}}ka, J., et al.\ The nature of light variations in magnetic hot stars.\ Contributions of the Astronomical Observatory Skalnate Pleso {\bf 48}, 170-174 (2018).
\bibitem{shavrina2010}        Shavrina, A.~V., et al.\ Spots structure and stratification of helium and silicon in the atmosphere of He-weak star HD 21699.\, \mnras {\bf 401}, 1882-1888 (2010).
\bibitem{glag2007}            Glagolevskij, Y.~V., \& Chuntonov, G.~A.\ Composite model for the magnetic field of HD 21699.\ Astrophysics {\bf 50}, 362-371 (2007).
\bibitem{cassisi2013}         Cassisi, S., \& Salaris, M.\ {\it Old Stellar Populations: How to Study the Fossil Record of Galaxy Formation.}\  (Wiley-VCH, 2013).
\bibitem{cantiello2019}       Cantiello, M., \& Braithwaite, J.\ Envelope Convection, Surface Magnetism, and Spots in A and Late B-type Stars.\, \apj {\bf 883}, 106 (2019).
\bibitem{cantiello2011}       Cantiello, M., \& Braithwaite, J.\ Magnetic spots on hot massive stars.\, \aap {\bf 534}, A140 (2011).
\bibitem{chris2012}           Moni Bidin, C., et al.\ Spectroscopy of horizontal branch stars in {\ensuremath{\omega}} Centauri{\ensuremath{\star}}.\, \aap {\bf 547}, A109 (2012).
\bibitem{schaefer2012}        Schaefer, B.~E.\ Astrophysics: Startling superflares.\, \nat {\bf 485}, 456-457 (2012).
\bibitem{schaefer2000}        Schaefer, B.~E., King, J.~R., \& Deliyannis, C.~P.\ Superflares on Ordinary Solar-Type Stars.\, \apj {\bf 529}, 1026-1030 (2000).
\bibitem{schaefer1989}        Schaefer, B.~E.\ Flashes from Normal Stars.\, \apj {\bf 337}, 927 (1989).
\bibitem{reed2014}            Reed, M.~D., et al.\ Analysis of the rich frequency spectrum of KIC 10670103 revealing the most slowly rotating subdwarf B star in the Kepler field.\, \mnras {\bf 440}, 3809-3824 (2014).
\bibitem{balona2015}          Balona, L.~A.\ Flare stars across the H-R diagram.\, \mnras {\bf 447}, 2714-2725 (2015).
\bibitem{fontaine2006}        Fontaine, G., Brassard, P., Charpinet, S., \& Chayer, P.\ The need for radiative levitation for understanding the properties of pulsating sdB stars.\, \memsai {\bf 77}, 49 (2006).
\bibitem{miller2020}    Miller   Bertolami,   M.~M.,    Battich,   T., C{\'o}rsico,  A.~H.,  Christensen-Dalsgaard,  J., \&  Althaus,  L.~G.\ Asteroseismic signatures  of the helium core  flash.\ Nature Astronomy {\bf 4}, 67-71 (2020).
%
\bibitem{landstreet2012}      Landstreet, J.~D., Bagnulo, S., Fossati, L., Jordan, S., \& O'Toole, S.~J.\ The magnetic fields of hot subdwarf stars.\, \aap {\bf 541}, A100 (2012).
\bibitem{bagnulo2015}         Bagnulo, S., Fossati, L., Landstreet, J.~D., \& Izzo, C.\ The FORS1 catalogue of stellar magnetic field measurements.\, \aap {\bf 583}, A115 (2015).
\bibitem{somers2020}          Somers, G., Cao, L., \& Pinsonneault, M.~H.\ The SPOTS Models: A Grid of Theoretical Stellar Evolution Tracks and Isochrones for Testing the Effects of Starspots on Structure and Colors.\, \apj {\bf 891}, 29 (2020).
\bibitem{saders2016}          van Saders, J.~L., et al.\ Weakened magnetic braking as the origin of anomalously rapid rotation in old field stars.\, \nat {\bf 529}, 181-184 (2016).
\bibitem{balona2019b}         Balona, L.~A.\ Evidence for spots on hot stars suggests major revision of stellar physics.\, \mnras {\bf 490}, 2112-2116 (2019).
\bibitem{balona2019a}         Balona, L.~A., et al.\ Rotational modulation in TESS B stars.\, \mnras {\bf 485}, 3457-3469 (2019).
\bibitem{dupuis2000}          Dupuis, J., Chayer, P., Vennes, S., Christian, D.~J., \& Kruk, J.~W.\ Adding More Mysteries to the DA White Dwarf GD 394.\, \apj {\bf 537}, 977-992 (2000).
\bibitem{kilic2015}           Kilic, M., et al.\ A Dark Spot on a Massive White Dwarf.\, \apjl {\bf 814}, L31 (2015).
\bibitem{koester1990}         Koester, D., \& Chanmugam, G.\ REVIEW: Physics of white dwarf stars.\ {\it Rep. on Prog. in Phys.}\/ {\bf 53}, 837-915 (1990).
\bibitem{reding2018}          Reding, J.~S., Hermes, J.~J., \& Clemens, J.~C.\ An Exploration of Spotted White Dwarfs from K2.\ In {\it Proceedings of the 21st European Workshop on White Dwarfs}, 1-6 (2018).
\bibitem{mestel1966}          Mestel, L.\ The magnetic field of a contracting gas cloud. I,Strict flux-freezing.\, \mnras {\bf 133}, 265 (1966).
\bibitem{ferrario2020}        Ferrario, L., Wickramasinghe, D.~T., \& Kawka, A.\ Magnetic fields in isolated and interacting white dwarfs.\ arXiv e-prints  arXiv:2001.10147 (2020).
\bibitem{grun1999}            Grundahl, F., Catelan, M., Landsman, W.~B., Stetson, P.~B., \& Andersen, M.~I.\ Hot Horizontal-Branch Stars: The Ubiquitous Nature of the ``Jump'' in Str{\"o}mgren u, Low Gravities, and the Role of Radiative Levitation of Metals.\, \apj {\bf 524}, 242-261 (1999).
\bibitem{pietru2017}          Pietrukowicz, P., et al.\ Blue large-amplitude pulsators as a new class of variable stars.\ \natast {\bf 1}, 0166 (2017).
%
\end{thebibliography}

\begin{thebibliography}{1}
\bibitem{fevre2003}        Le F{\`e}vre, O., et al.\ Commissioning and performances of the VLT-VIMOS instrument.\ \procspie {\bf 4841}, 1670-1681 (2003).
\bibitem{kuijken2011}      Kuijken, K.\ OmegaCAM: ESO's Newest Imager.\ {\it The Messenger}\/ {\bf 146}, 8-11 (2011).
\bibitem{pbs1987}          Stetson, P.~B.\ DAOPHOT: A Computer Program for Crowded-Field Stellar Photometry.\ \pasp {\bf 99}, 191 (1987).
\bibitem{alard1998}        Alard, C., \& Lupton, R.~H.\ A Method for Optimal Image Subtraction.\ \apj {\bf 503}, 325-331 (1998).
\bibitem{montalto2007}     Montalto, M., et al.\ A new search for planet transits in NGC~6791.\ \aap {\bf 470}, 1137-1156 (2007).
\bibitem{bjd}              Eastman, J., Siverd, R., \& Gaudi, B.~S.\ Achieving Better Than 1 Minute Accuracy in the Heliocentric and Barycentric Julian Dates.\ \pasp {\bf 122}, 935 (2010).
\bibitem{vartool}          Hartman, J.~D., \& Bakos, G. {\'A}.\ VARTOOLS: A program for analyzing astronomical time-series data.\ {\it Astronomy and Computing}\/ {\bf 17}, 1-72 (2016).
\bibitem{schwa1}           Schwarzenberg-Czerny, A.\ Fast and Statistically Optimal Period Search in Uneven Sampled Observations.\ \apjl {\bf 460}, L107 (1996).
\bibitem{schwa2}           Schwarzenberg-Czerny, A., \& Beaulieu, J.-P.\ Efficient analysis in planet transit surveys.\ \mnras {\bf 365}, 165-170 (2006).
\bibitem{kaluzny2004}      Kaluzny, J., et al.\ Cluster AgeS ExperimentCatalog of variable stars in the globular cluster {\ensuremath{\omega}} Centauri.\ \aap {\bf 424}, 1101-1110 (2004).
\bibitem{kaluzny2009}      Kaluzny, J., \& Thompson, I.~B.\ Variable Stars in the Globular Cluster NGC 6752.\ \actaa {\bf 59}, 273-289 (2009).
\bibitem{rozy2017}         Rozyczka, M., et al.\ The Cluster AgeS Experiment (CASE). Variable Stars in the Field of the Globular Cluster M22.\ \actaa {\bf 67}, 203-224 (2017).
\bibitem{rozy2020}         Rozyczka, M., et al.\ The Cluster AgeS Experiment (CASE). Variable stars in the field of the globular cluster M10.\ arXiv e-prints  arXiv:2001.01529 (2020).
\bibitem{milone2015}       Milone, A.~P., et al.\ The Hubble Space Telescope UV Legacy Survey of Galactic Globular Clusters. III. A Quintuple Stellar Population in NGC 2808.\ \apj {\bf 808}, 51 (2015).
\bibitem{ferraro1998}      Ferraro, F.~R., Paltrinieri, B., Fusi Pecci, F., Rood, R.~T., \& Dorman, B.\ Multimodal Distributions along the Horizontal Branch.\ \apj {\bf 500}, 311-319 (1998).
\bibitem{rolly2000}        Bedin, L.~R., et al.\ The anomalous Galactic globular cluster NGC 2808. Mosaic wide-field multi-band photometry.\ \aap {\bf 363}, 159-173 (2000).
\bibitem{pietrinferni2013} Pietrinferni, A., Cassisi, S., Salaris, M., \& Hidalgo, S.\ The BaSTI Stellar Evolution Database: models for extremely metal-poor and super-metal-rich stellar populations.\ \aap {\bf 558}, A46 (2013).
\bibitem{nardiello2018}    Nardiello, D., et al.\ The Hubble Space Telescope UV Legacy Survey of Galactic Globular Clusters - XVII. Public Catalogue Release.\ \mnras {\bf 481}, 3382-3393 (2018).
\bibitem{cool2013}         Cool, A.~M., et al.\ HST/ACS Imaging of Omega Centauri: Optical Counterparts of Chandra X-Ray Sources.\ \apj {\bf 763}, 126 (2013).
\bibitem{chris2007}        Moni Bidin, C., Moehler, S., Piotto, G., Momany, Y., \& Recio-Blanco, A.\ Spectroscopy of horizontal branch stars in NGC~6752. Anomalous results on atmospheric parameters and masses.\ \aap {\bf 474}, 505-514 (2007).
\bibitem{dorazi2015}       D'Orazi, V., et al.\ Lithium abundances in globular cluster giants: NGC 1904, NGC 2808, and NGC 362.\ \mnras {\bf 449}, 4038-4047 (2015).
\bibitem{coelho2014}       Coelho, P.~R.~T.\ A new library of theoretical stellar spectra with scaled-solar and {\ensuremath{\alpha}}-enhanced mixtures.\ \mnras {\bf 440}, 1027-1043 (2014).
\bibitem{jones2017}        Jones, D., \& Boffin, H.~M.~J.\ Binary stars as the key to understanding planetary nebulae.\ \natast {\bf 1}, 0117 (2017).
\bibitem{prsa2016}         Pr{\v{s}}a, A., et al.\ Physics Of Eclipsing Binaries. II. Toward the Increased Model Fidelity.\ \apjs {\bf 227}, 29 (2016).
\bibitem{bertelli2008}     Bertelli, G., Girardi, L., Marigo, P., \& Nasi, E.\ Scaled solar tracks and isochrones in a large region of the Z-Y plane. I. From the ZAMS to the TP-AGB end for 0.15-2.5 M\ensuremath{_\odot} stars.\ \aap {\bf 484}, 815-830 (2008).
\bibitem{hillwig2016}      Hillwig, T.~C., et al.\ Observational Confirmation of a Link Between Common Envelope Binary Interaction and Planetary Nebula Shaping.\ \apj {\bf 832}, 125 (2016).
\bibitem{gilli2000}        Gilliland, R.~L., et al.\ A Lack of Planets in 47 Tucanae from a Hubble Space Telescope Search.\ \apjl {\bf 545}, L47-L51 (2000).
\bibitem{valerio2012}      Nascimbeni, V., Bedin, L.~R., Piotto, G., De Marchi, F., \& Rich, R.~M.\ An HST search for planets in the lower main sequence of the globular cluster NGC 6397.\ \aap {\bf 541}, A144 (2012).
\bibitem{wallace2020}      Wallace, J.~J., Hartman, J.~D., \& Bakos, G. {\'A}.\ A Search for Transiting Planets in the Globular Cluster M4 with K2: Candidates and Occurrence Limits.\ \aj {\bf 159}, 106 (2020).
\bibitem{santander2015}    Santander-Garc{\'\i}a, M., et al.\ The double-degenerate, super-Chandrasekhar nucleus of the planetary nebula Henize 2-428.\ \nat {\bf 519}, 63-65 (2015).
\bibitem{vos2018}          Vos, J., N{\'e}meth, P., Vu{\v{c}}kovi{\'c}, M., {\O}stensen, R., \& Parsons, S.\ Composite hot subdwarf binaries - I. The spectroscopically confirmed sdB sample.\ \mnras {\bf 473}, 693-709 (2018).
\end{thebibliography}
\end{document}